\documentclass[twocolumn,showpacs,preprintnumbers,amsmath,amssymb]{revtex4}
\usepackage{tabularx,graphicx}

\usepackage{color}
\usepackage{hyperref}
\hypersetup{
    colorlinks=true,
    linkcolor=blue,
    filecolor=blue,      
    urlcolor=blue,
}

\begin{document}

\newcommand{\beq}{\begin{equation}}
\newcommand{\eeq}{\end{equation}}
\newcommand{\beqn}{\begin{eqnarray}}
\newcommand{\eeqn}{\end{eqnarray}}
\newcommand{\bmath}{\begin{subequations}}
\newcommand{\emath}{\end{subequations}}
\newcommand{\bra}[1]{\langle #1|}
\newcommand{\ket}[1]{|#1\rangle}

\title{How Alfven's theorem explains the Meissner effect}
\author{J. E. Hirsch }
\address{Department of Physics, University of California, San Diego,
La Jolla, CA 92093-0319}

\begin{abstract} 
Alfven's theorem states that in a perfectly conducting fluid magnetic field lines move with the fluid without dissipation.
When a  metal becomes superconducting in the presence of a magnetic field, magnetic field lines move from the interior to the surface (Meissner effect) in a reversible way. This indicates that a perfectly conducting fluid is flowing outward.
We point this out and  show  that this fluid carries neither charge nor mass, but carries  {\it effective mass}. This implies that the effective mass of carriers
is lowered when a system goes from the normal to the superconducting state, which agrees with the prediction of the
unconventional    theory of hole superconductivity and with optical experiments in some superconducting materials. The 
60-year old  conventional understanding of the Meissner effect ignores Alfven's theorem and for that reason we argue that it does not provide a valid understanding  of real superconductors.  \end{abstract}
\pacs{}
\maketitle

\section{introduction}

When a conducting fluid moves, magnetic field lines tend to move with the fluid, as a consequence of
Faraday's law \cite{davidson}. If the fluid is perfectly conducting, the lines are `frozen' in the fluid. That is known as `Alfven's theorem' \cite{alfventheorem}. No dissipation occurs when
a perfectly  conducting  fluid together with   magnetic field lines move. If the fluid is not  perfectly conducting, there will  be  
relative motion of magnetic field lines with respect to the fluid and   Joule heat will be dissipated \cite{davidson}.
Even for non-perfectly conducting fluids, as P. H. Roberts points out \cite{roberts},
{\it ``Alfven's theorem is also helpful
in attacking the problem of inferring unobservable fluid motions from
observed magnetic field behavior''}. For example, measurements of magnetic field variations near
one of Jupiter's moons demonstrated the existence of an unobservable conducting fluid below its surface \cite{kivelsonmg}. 
This paper is based on Roberts' principle.

In the transition from normal metal to superconductor in the presence of a magnetic field, magnetic field lines move out of the interior of the system. This is called  the Meissner effect. The transition is thermodynamically reversible, i.e. it occurs without dissipation under ideal conditions. 
In both the normal and the superconducting states of the metal there are mobile electric charges, which certainly
qualify as a conducting fluid. 
Thus it is logical to infer that the motion of magnetic field lines in the normal-superconductor transition  is associated with the motion of  charges, specifically  that the motion of magnetic field lines $reflects$ the motion of charges.
In this paper we propose that this is indeed the case, and explain what  the nature of this conducting  fluid is and {\it what this
fluid motion carries with it} in addition to the magnetic field. 

Instead, the conventional (BCS) theory of superconductivity \cite{tinkham} says that the outward motion of magnetic field lines in the normal-superconductor transition is determined by quantum mechanics and energetics and is $not$ associated with the outward motion of any charges. 
We will argue that this is incorrect.

In earlier work we have used related concepts to explain the physics of the Meissner effect based on the theory of
hole superconductivity proposed to describe all superconducting materials \cite{holesc,libro}.  This will be discussed
later in the paper.
 
        \begin{figure} []
 \resizebox{8.5cm}{!}{\includegraphics[width=6cm]{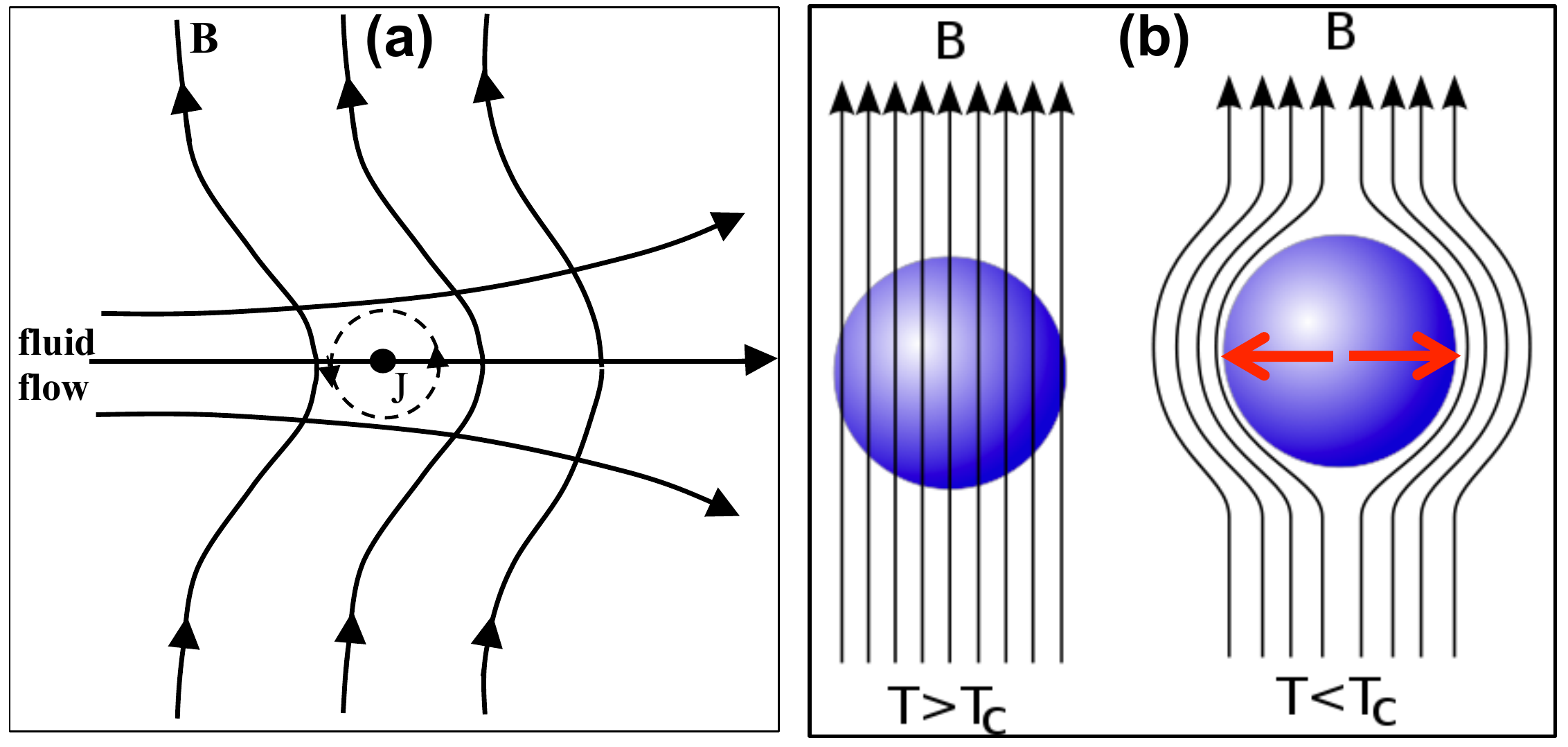}} 
 \caption { The left panel shows an example illustrating Alfven's theorem for a conducting fluid \cite{davidson}.
 Fluid flow across magnetic field lines causes the field lines to bow out.  The right panel
 shows the Meissner effect in a superconductor. The red arrows show the hypothesized motion of
 fluid, by analogy to the left panel. }
 \label{figure1}
 \end{figure} 
 
 \section{alfven's theorem}
 
 When a conducting fluid moves with velocity $\vec{u}$ in the presence of electric and magnetic fields
 $\vec{E}$ and $\vec{B}$,
 electromagnetism dictates that an electric current density \cite{davidson}
 \beq
 \vec{J}=\sigma (\vec{E}+\frac{1}{c}\vec{u}\times\vec{B}) 
 \eeq
 exists, where $\sigma$ is the electrical conductivity of the fluid. In particular, for a perfectly conducting
 fluid $\sigma=\infty$ and 
 \beq
 \vec{E}=-\frac{1}{c}\vec{u}\times\vec{B} .
 \eeq
 Fig. 1 shows in the left panel qualitatively
 how this leads to Alfven's theorem. The horizontal motion of the fluid
 generates a current $J$ pointing out of the paper which generates a counterclockwise
 magnetic field indicated by the dashed circle, which added to the original magnetic field
 gives curvature to the magnetic field lines that were originally straight. Thus, the magnetic
 field lines bend in the direction of the fluid motion. 
 
 Analogously we suggest in this paper that the motion of magnetic field lines
 in the right panel of Fig. 1 is associated with  motion of a conducting  fluid as indicated by the red arrows.
 
Using Faraday's and Ampere's laws,
\beq
\vec{\nabla}\times\vec{E}=-\frac{1}{c}\frac{\partial \vec{B}}{\partial t},
\eeq
\beq
\vec{\nabla}\times\vec{B}=\frac{4\pi}{c}\vec{J}
\eeq
Eq. (1) yields
\beq
\frac{\partial\vec{B}}{\partial t}=\vec{\nabla}\times(\vec{u}\times\vec{B})+\frac{c^2}{4\pi\sigma}\nabla^2\vec{B}  
\eeq
and in particular for a perfectly conducting fluid  
\beq
\frac{\partial\vec{B}}{\partial t}=\vec{\nabla}\times(\vec{u}\times\vec{B}) .
\eeq
 Eq. (6) implies  that   magnetic field lines are frozen into the fluid. The proof is given  in Appendix A. This implies
that for a perfectly conducting fluid  outward motion of field lines is necessarily associated with outward motion of the fluid.

For generality, we could assume that in addition to the current given by Eq. (1) there is 
a  `quantum supercurrent' $\vec{J}_s$ generated by an unknown quantum mechanism
provided by BCS or another microscopic theory:
 \beq
 \vec{J}=\sigma (\vec{E}+\frac{1}{c}\vec{u}\times\vec{B})+\vec{J}_s .
 \eeq
 Instead of Eq. (5) we would  obtain from Eq. (7)
\beq
\frac{\partial\vec{B}}{\partial t}=\vec{\nabla}\times(\vec{u}\times\vec{B})+\frac{c^2}{4\pi\sigma}\nabla^2\vec{B}
+\frac{c}{\sigma}\vec{\nabla}\times\vec{J}_s .
\eeq

Consider a long metallic cylinder initially in the normal state with uniform magnetic field in the $\hat{z}$ direction. In cylindrical coordinates and assuming translational invariance in the
$\hat{z}$ and $\hat{\theta}$ (azimuthal) directions Eq. (8) yields for the time evolution of the magnetic field
$\vec{B}=B(r,t)\hat{z}$
\beq
\frac{\partial B(r,t)}{\partial t}=-\frac{1}{r}\frac{\partial (ru_r B)}{\partial r} 
+\frac{c}{\sigma}  \frac{1}{r}    \frac{\partial}{\partial r}(rJ_{s\theta}) .
\eeq
Note that the last term, the contribution of the `quantum supercurrent' to the time evolution of the magnetic field, decreases as $\sigma$ increases.
Thus it is natural to conclude that for large $\sigma$ at least the time evolution of  the magnetic field is
dominated by the first term in Eq. (9), which requires $radial$ motion of the fluid, 
$u_r\neq 0$, i.e. {\it motion of the conducting fluid in direction perpendicular to the field lines}.

Within the conventional theory  of superconductivity \cite{tinkham}  $u_r=0$ and the expulsion of magnetic field has
to be explained solely  by the last term in Eq. (9).  The explanation has to be valid for any value of $\sigma$, since normal metals of widely varying
conductivities expel magnetic fields when they become superconducting. How this happens within the conventional theory  has not been explained in the
literature.

Instead, in this paper we will assume that the last term in Eq. (9) doesn't exist and   explain the Meissner effect
in a natural way 
through the  outward motion of a perfectly conducting fluid.

\section{the puzzle}
A perfectly conducting fluid moving from the interior to the surface when a normal metal becomes superconducting
would satisfy Eq. (6) and as a consequence, as shown in   Appendix A,
would carry the magnetic field lines with it and explain the Meissner effect. 
However, there are obvious problems with this  explanation:

(1) If the fluid is charged, this motion would result in an inhomogeneous charge distribution, costing an
enormous electrostatic potential energy. So this cannot happen. 

(2) Even if the  fluid is charge neutral,   like a neutral plasma composed of electrons and ions with equal and opposite charge
densities, outward motion would be associated with outward mass flow,
generating an enormous mass imbalance. This cannot happen. Plasmas cannot expel magnetic fields by outward motion.

(3) Furthermore, in a solid the positive  ions cannot move a finite distance. The only mobile charges  are electrons.

So in order to explain the Meissner effect using Alfven's theorem we need  to identify a charge-neutral mass-neutral electricity-conducting fluid that
moves from the interior to the surface in the process of the metal becoming superconducting, without dissipation.

And this poses an additional question: if this fluid  carries neither charge nor mass, what does it carry?

The next section provides the answers.

\section{the answers}

Charge carriers in electronic energy bands can be electrons or holes \cite{am}. We will need both to explain how magnetic  flux is expelled.

Consider a long metallic cylinder of radius $R$,   of a material that is  a type I superconductor,  in a uniform applied magnetic field $H=H_c(T)$ parallel to its
axis, where $H_c(T)$ is the critical magnetic field at temperature $T$ \cite{tinkham}, that is initially at temperature higher than $T$. 
When the system is cooled to temperature $T$ it will become superconducting and expel the magnetic field to
a surface layer of thickness $\lambda_L$, the London penetration depth at that temperature, 
typically hundreds of $\AA$, given by \cite{tinkham}
\beq
\frac{1}{\lambda_L^2}=\frac{4\pi n_s e^2}{m^* c^2}
\eeq
with $n_s$ the density of superfluid carriers and $m^*$ their effective mass.

Assume that the transition proceeds as follows. Initially, in  a central core of radius $r_c$ a
perfectly conducting  fluid of $n_s$ electrons and $n_s$ holes per unit volume forms, both carriers with  effective
mass $m^*$, with 
$r_c$ given by
\beq
r_c=\sqrt{2R\lambda_L}
\eeq  
as shown in the left panel of Fig. 2.
Then, assume this fluid flows radially outward until it reaches the surface. Assuming it is incompressible, it will at the end
occupy an annulus of thickness  $\lambda_L$ adjacent to the surface, since
\beq
\pi r_c^2=2\pi R \lambda_L .
\eeq

        \begin{figure} []
 \resizebox{8.5cm}{!}{\includegraphics[width=6cm]{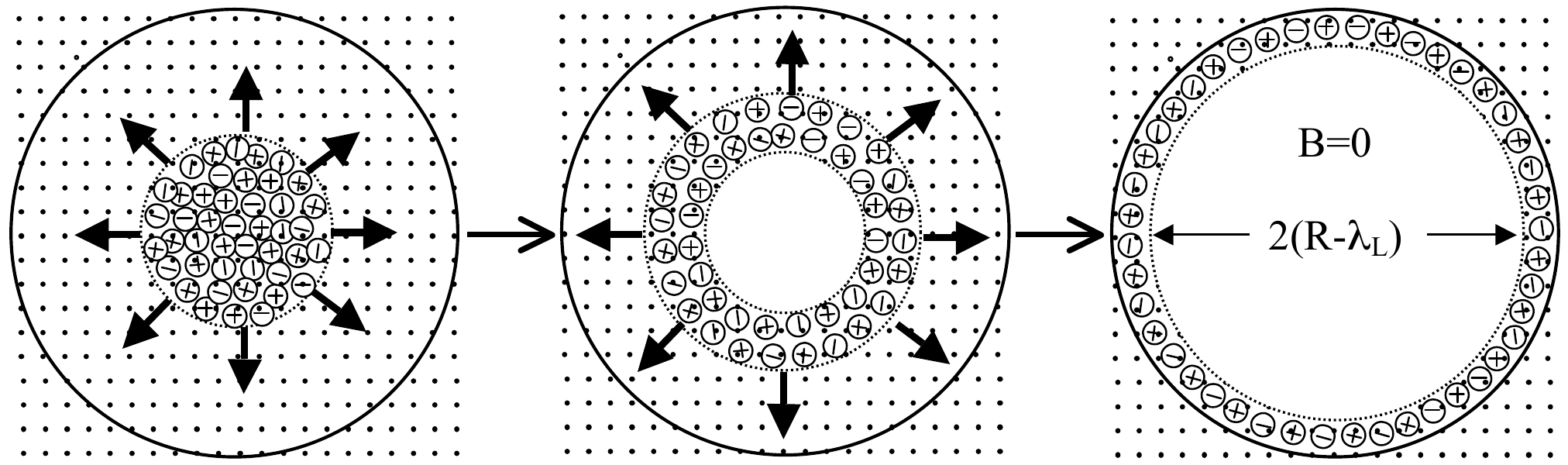}} 
 \caption { Simple model for the Meissner effect in a cylinder (top view).  A perfectly conducting fluid of electrons and holes occupies initially the central region  (of radius $r_c$) of a cylinder of radius $R$ (left panel)
 and flows to the surface where it occupies a ring of thickness  $\lambda_L$. Points indicate magnetic field pointing out of the paper, initially uniformly
 distributed across the cylinder cross section. }
 \label{figure1}
 \end{figure} 

Because of Alfven's theorem, the magnetic field lines that were initially in the region $r\leq r_c$  flow out with the
fluid. No magnetic field line can cross either the inner or  the outer boundary of this fluid, therefore the magnetic field lines that
were outside initially are pushed further out as the fluid moves out, and in the interior no magnetic field ever exists. The end result is what is shown on the right
panel of Fig. 2: no magnetic field in the region $r<R-\lambda_L$. The magnetic field has been expelled from the interior and remains only in 
a surface layer of thickness $\lambda_L$, as occurs in the Meissner effect.

There is however a  small difference. Frozen field lines would imply that in the final state the magnetic field is uniform   in the region $R-\lambda_L<r<R$ and drops discontinuously to zero at $r=R-\lambda_L$. This is not so in the Meissner effect, rather  the magnetic field near the surface is given by (to lowest order in $\lambda_L/R$)
\beq
H(r)=H_ce^{(r-R)/\lambda_L} .
\eeq
The reason for the difference is that in assuming that Eq. (6) is valid at all times we are ignoring transient effects and the inertia of
charge carriers.
This is a minor difference, in particular the magnetic field flux through the region $r\leq R$ is the same for Eq. (13) 
as it is  for 
a uniform $H_c$ between  $R-\lambda_L$ and $R$.

The fluid that flowed out is charge neutral, by assumption, so no charge imbalance results from this process.
Furthermore {\it no mass imbalance} results from this process either. To understand this one has to
remember that `holes' are not real particles, they are a theoretical construct \cite{am}. When holes flow in 
a given direction, physical mass is flowing in the opposite direction. This is illustrated in figure 3.
So the process that we envision shown in Fig. 2 would result if we have conduction in two bands,
  one  close to empty and the other one close to full,  with the same density of electrons and holes. 
  This is depicted in Fig.  4.
  
  But if neither charge nor mass flowed out, what quantity is being transported out in the process shown
  in Fig. 2?
  
       \begin{figure} [t]
 \resizebox{8.5cm}{!}{\includegraphics[width=6cm]{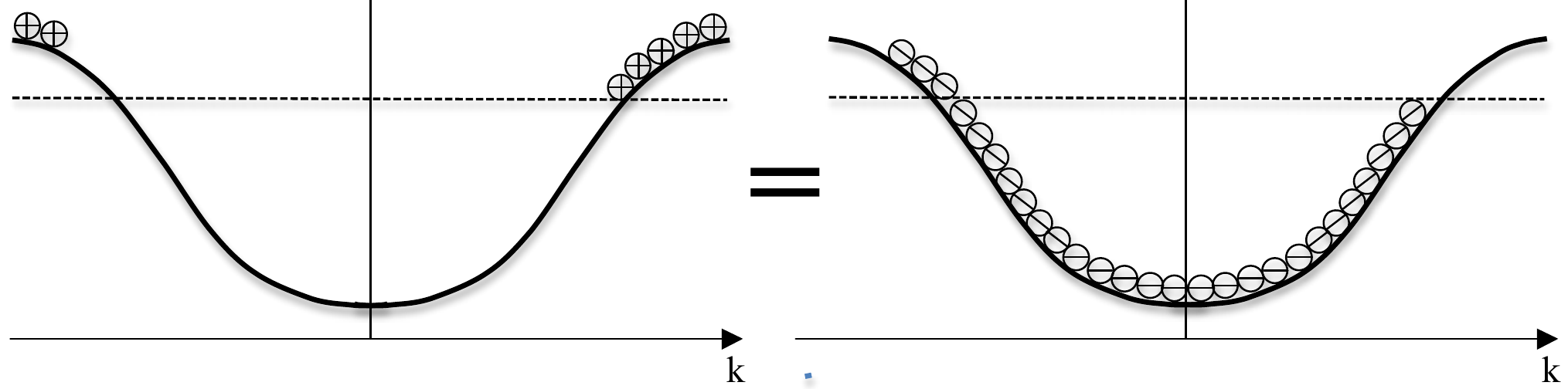}} 
 \caption { Holes flowing in the positive $k$ direction (left panel) corresponds to electrons flowing in the
 negative $k$ direction (right panel).}
 \label{figure1}
 \end{figure} 
           \begin{figure} [t]
 \resizebox{8.5cm}{!}{\includegraphics[width=6cm]{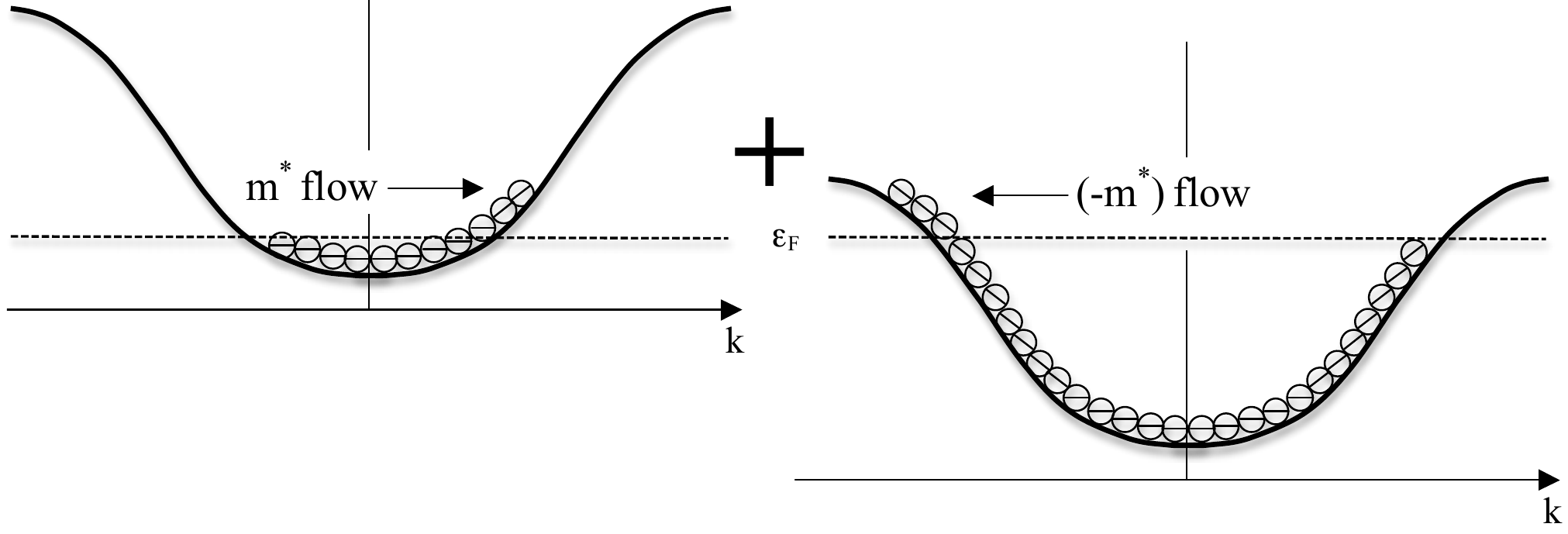}} 
 \caption { The left panel shows electrons flowing out (positive $k$ direction) carrying $out$  positive effective mass.
 The right panel shows holes flowing out (see left panel of Fig. 3), carrying $in$ (negative $k$ direction)  $negative$ effective mass,
 which is equivalent to carry out positive effective mass also. Both of these flows occur in Fig. 2. $\epsilon_F$ is the Fermi energy.}
 \label{figure1}
 \end{figure} 
 
  The answer is, {\it effective mass}. The effective mass {\it of electrons} is given by the curvature of the
  energy bands in Fig. 4.
Having holes with positive charge and positive effective mass flowing out is
  equivalent to having electrons with negative charge and negative effective mass flowing in, as shown in Fig. 3.
  So the electron band carriers carry out positive effective mass, and the hole band carriers carry $in$ negative 
  effective mass,  which is equivalent to  also carrying  out positive effective mass.   
  This implies that there is a net {\it outflow of effective mass} in the process where a metal going superconducting
  expels magnetic field. In addition to expelling magnetic field, the system expels effective mass
  
As a consequence,  in the process of a metal going superconducting the effective mass of the carriers in the
  system decreases. We discuss this further in Sect. VII.

           \begin{figure} [t]
 \resizebox{6.5cm}{!}{\includegraphics[width=6cm]{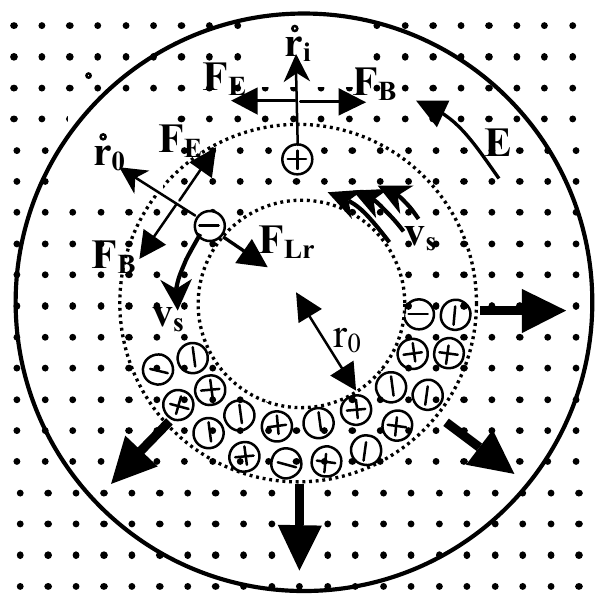}} 
 \caption {  Expulsion of magnetic field (dots) through motion of perfectly conducting fluid.
 The charge distribution has azimuthal symmetry but only some of the carriers are shown for clarity.
 Both electrons and holes move radially out with speed $\dot{r}_0$. In addition electrons have azimuthal speed $v_s$ that nullifies the
 magnetic field in the interior. The electric and magnetic Lorentz forces are balanced in the azimuthal direction for
 both electrons and holes. For electrons there is also a radial Lorentz force $F_{Lr}$ that is balanced by
 quantum pressure (see text).  }
 \label{figure1}
 \end{figure}

 \section{Kinetics  of the fluid motion}
Equation (6) guarantees that no magnetic field lines can cross the boundaries of our annulus of perfectly conducting fluid as it moves outward, as shown
in   Appendix A.   Let us consider the current distribution. 
The details of the current distribution will depend on the initial conditions. We assume initial conditions so that it is only the electrons that
have azimuthal velocity. Figure 5 shows an intermediate state in the process.

$r_0(t)$ is the inner  radius of the annulus of fluid that is moving outward with speed $\dot{r}_0$. The fluid velocity field
is   given by
\beq
\vec{u}(r)=\dot{r}_0 \frac{r_0}{r} \hat{r}.
\eeq
We assume that $\dot{r}_0$ is of order of the speed at which superconductors expel magnetic fields in experiments \cite{faber},
typically $mm/s$, hence $much$ smaller than the speed of light.
The magnetic field is zero for $r\leq r_0$ according to Alfven's theorem, and it is given by $H_c$ for $r>>r_0$. It cannot go to zero discontinuously at $r_0$ unless the current density at $r=r_0$ is infinite. So we assume it goes continuously to
zero in a region of thickness $\lambda$ adjacent to the surface where current flows. It is natural to assume that
the decay is exponential, so we assume the form
\beq
\vec{H}(r)=H_c(1-e^{(r_0-r)/\lambda})\hat{z} .
\eeq
In the next section  we will  show that the decay is indeed exponential and that $\lambda=\lambda_L$, with $\lambda_L$ the London penetration depth given by Eq. (10).

Using Faraday's law and the fact that the magnetic field is zero in the deep interior we obtain for the Faraday electric field
\beq
\vec{E}(r)=\frac{\dot{r}_0}{c}\frac{r_0}{r}H_c(1- e^{(r_0-r)/\lambda})\hat{\theta}
\eeq
(to lowest order in $\lambda/r$). The azimuthal velocity for the electrons in the annulus  is
\beq
\vec{v}_s(r)=-\frac{c}{4\pi en_s\lambda}H_c e^{(r_0-r)/\lambda} \hat{\theta}
\eeq
giving rise to azimuthal current density
\beq
\vec{J}=n_s e \vec{v}_s=-\frac{c}{4\pi\lambda } H_c  e^{(r_0-r)/ \lambda} \hat{\theta} .
\eeq
The current density Eq. (18) satisfies   Ampere's law
\beq
\vec{\nabla}\times\vec{H}=\frac{4\pi}{c}\vec{J} .
\eeq
From Eq. (15)
\beq
\frac{\partial \vec{H}}{\partial t}=-\frac{\dot{r}_0}{\lambda} \vec{H} .
\eeq
and Eq. (6) is satisfied to lowest order in $\lambda_L/r$, with $\vec{u}$ given by Eq. (14).  The electric field,  magnetic field and 
fluid velocity  Eqs. (15), (13) and (14) are
related by the condition
\beq
\vec{E}=-\frac{1}{c}\vec{u}\times\vec{H}
\eeq
in agreement with Eq. (2).  

The Lorentz force
\beq
\vec{F}_L=q(\vec{E}+\frac{1}{c}\vec{v}\times\vec{B}) \equiv \vec{F}_E+\vec{F}_B
\eeq
 {\it in the azimuthal direction}
is zero for both electrons and holes with $v_r=u(r)$, as shown schematically  in Fig. 5.
For electrons there is also a Lorentz force in the radial direction
\beq
F_{Lr}=\frac{e}{c} v_s H \hat{r}=-\frac{1}{4\pi n_s \lambda} e^{2(r_0-r)/\lambda}H_c^2 \hat{r}
\eeq
so in order for this fluid to move outward there has to be an outward force that compensates the inward force Eq. (23). That outward
force $F_r=-F_{Lr}$ (per unit area)  is called ``Meissner pressure'' and it arises from the difference in energy between normal and superconducting states. 
From Eq. (23) we obtain 
the work done by $F_r$ per unit area per unit time:
\beq
\int^\infty _{r_0} dr F_r v_r n_s=\frac{H_c^2}{8\pi}\dot{r}_0
\eeq
which is the rate of change of magnetic energy per unit area as the phase boundary moves. This energy is provided
by the condensation energy of the superconductor. 

   \section{Dynamics of the fluid  motion}
    Here we show that the magnetic and velocity fields  discussed in Sect. V indeed have exponential dependence
    on $r$ as assumed and decay   length $\lambda$ given by the London penetration depth   $\lambda_L$, Eq. (10).

    The equation of motion for electrons of effective mass $m^*$ in electric and magnetic fields in a perfectly conducting fluid  is
    \beq
    \frac{d\vec{v}}{dt}=\frac{e}{m^*}\vec{E}+\frac{e}{m^*c}\vec{v}\times \vec{H}
    \eeq
    Using the relation between total and partial time derivatives, Eq. (25) becomes
    \beq
        \frac{\partial \vec{v}}{\partial t}+\vec{\nabla} (\frac{v^2}{2})-\vec{v}\times(\vec{\nabla}\times\vec{v}) =\frac{e}{m^*}\vec{E}+\frac{e}{m^*c}\vec{v}\times \vec{H} .
        \eeq
        In cylindrical coordinates, the velocity field is
        \beq
        \vec{v}(r,t)=\vec{v}_\theta (r,t)\hat{\theta }+\frac{r_0}{r}\dot{r}_0\hat{r}
        \eeq
        so for the azimuthal direction Eq. (26) yields
        \beq
        \frac{\partial v_\theta}{\partial t}+\dot{r}_0\frac{r_0}{r^2}\frac{\partial }{\partial r }(r v_\theta)=\frac{e}{m^*}E+\frac{e}{m^*c}\dot{r}_0\frac{r_0}{r}H .
        \eeq
        
        On the other hand, by taking the curl on both sides of Eq. (26) we find
        \beq
        \frac{\partial \vec{w}}{\partial t}=\vec{\nabla}\times [\vec{v} \times\vec{w}]
        \eeq
        with
        \bmath
        \beq
        \vec{w}\equiv \vec{\nabla}\times\vec{v}+\frac{e}{m^*c} \vec{H} ,
        \eeq
        \beq
        \vec{w}=w(r,t)\hat{r} ,
        \eeq
        \emath
        and from Eq. (30a)
        \beq
        w(r,t)=\frac{1}{r}\frac{\partial}{\partial r}(r v_\theta)+\frac{e}{m^*c} H(r,t).
        \eeq
        In cylindrical coordinates Eq. (29) is
        \beq
        \frac{\partial w}{\partial t}=-\frac{r_0}{r}\dot{r}_0 \frac{\partial w}{\partial r}
        \eeq
        that is satisfied by
        \beq
        w(r,t)=g(r-\frac{r_0}{r}\dot{r}_0 t) .
        \eeq
        with g an arbitrary function. Now at $t=0$ we have from Eq. (31)
        \beq
        w(r,t=0)=\frac{e}{m^*c} H_c
        \eeq
        for all $r$, since the fluid has not started to move. Therefore, from Eq. (33)
        \beq
             w(r,0)=g(r )=\frac{e}{m^*c} H_c
             \eeq
             for all $r$. Therefore, $w$ is simply given by
             \beq
             w(r,t)=\frac{e}{m^*c} H_c
             \eeq
             and from Eq. (31)
                 \beq
\frac{1}{r}\frac{\partial}{\partial r}(r v_\theta)=\frac{e}{m^*c} (H_c-H(r,t)).
        \eeq
        We now replace Eq. (37) in the equation of motion (28) and obtain
        \beq
        \frac{\partial v_\theta(r,t)}{\partial t}+\dot{r}_0\frac{r_0}{r}\frac{e}{m^*c} H_c= \frac{e}{m^*}E(r,t).
        \eeq
        Now from Ampere-Maxwell's law
        \beq
        \vec{\nabla}\times\vec{H}=\frac{4\pi}{c}\vec{J}+\frac{1}{c}\frac{\partial \vec{E}}{\partial t}
        \eeq
        using that
        \beq
        J_\theta=n_s e v_\theta
        \eeq
        Eq. (39) yields
        \beq
\frac{\partial E}{\partial t}=-c\frac{\partial H}{\partial r}-4\pi n_s e v_\theta .
\eeq
Taking the time derivative of Eq. (38) and using (41)
\beq
\frac{\partial ^2 v_\theta}{\partial t^2}=-\frac{ec}{m^*}\frac{\partial H}{\partial r}-\frac{4\pi n_s e^2}{m^*} v_\theta .
\eeq
             Taking the space derivative of Eq. (37)
 \beq
 \frac{\partial H}{\partial r}=-\frac{m^*c}{e}\frac{\partial}{\partial r}(\frac{1}{r}\frac{\partial}{\partial r}(r v_\theta))
 \eeq
 and replacing (43) in (42)
 \beq
\frac{1}{c^2}\frac{\partial ^2 v_\theta}{\partial t^2}= \frac{\partial}{\partial r}(\frac{1}{r}\frac{\partial}{\partial r}(r v_\theta))-\frac{4\pi n_s e^2}{m^*c^2} v_\theta .
\eeq
Eq. (44) describes the full time dependence of the process discussed in Sect. V including the initial transient
when the fluid starts to move and the azimuthal current is established. The initial conditions are
\bmath
\beq
v_\theta(r,t=0)=0 ,
\eeq
\beq
\frac{\partial v_\theta(r>0,t)}{\partial t})_{t=0}=0 ,
\eeq
\beq
\frac{\partial v_\theta(r=0,t)}{\partial t})_{t=0}=\frac{e}{m^*c} \dot{r}_0 H_c .
\eeq
\emath

 Now in Sect. V we assumed for the azimuthal velocity in the steady state situation
 \beq
v_\theta(r,t)=-\frac{c}{4\pi en_s\lambda}H_c e^{(\dot{r}_0t-r)/\lambda} 
\eeq
Its second time derivative is
\beq
\frac{1}{c^2}\frac{\partial ^2 v_\theta}{\partial t^2}=(\frac{\dot{r}_0}{c})^2\frac{v_\theta}{\lambda^2} .
\eeq 
Since we assume $\dot{r}_0<<c$, we conclude that  Eq. (47) is completely negligible and hence that after an initial transient where the velocity field is established, the left-hand side
of Eq. (44) is completely negligible. In steady state then Eq. (44) yields
\beq
   \frac{\partial}{\partial r}(\frac{1}{r}\frac{\partial}{\partial r}(r v_\theta))-\frac{1}{\lambda_L^2} v_\theta =0
   \eeq    
   with
   \beq
   \frac{1}{\lambda_L^2}=\frac{4\pi n_s e^2}{m^*c^2}  .
   \eeq
   the same as Eq. (10) for superconductors.

        The exact solution of Eq. (48) is simply obtained in terms of Bessel functions \cite{laue}. To lowest order in $\lambda_L/r$ it is 
\beq
v_\theta=Ce^{-r/\lambda_L}
\eeq
where C is independent of $r$. To find $C$, we use the fact that except for the initial transient we can ignore the Maxwell term in
Ampere-Maxwell's law Eq. (39), hence from Eq. (41)
\beq
v_\theta=-\frac{c}{4\pi n_s e } \frac{\partial H}{\partial r}
\eeq
and replacing in Eq. (48) and using Eq. (49)
\beq
   \frac{\partial}{\partial r}\frac{1}{r}\frac{\partial}{\partial r}(r H)-\frac{1}{\lambda_L^2} (H-H_c)=0
   \eeq    
   so that  $H-H_c$ and $v_\theta$ obey the same equation. To lowest order in $\lambda_L/r$ again the solution is
   \beq
   H(r)=H_c-C'e^{-r/\lambda_L} .
   \eeq
   \newline
Now  we use the condition $H(r=r_0)=0$ to get
   \beq
   C'=e^{r_0/\lambda_L}H_c
   \eeq
   hence 
   \beq
   H(r)=H_c(1-e^{(r_0-r)/\lambda_L})
   \eeq
   the same as Eq. (15). Replacing Eq. (55) in Eq. (51) we finally obtain
   \beq
   v_\theta=-\frac{c}{4\pi n_s e\lambda_L}H_ce^{(r_0-r)/\lambda_L}
   \eeq
   i.e. the same as Eq. (17), with $\lambda=\lambda_L$.
   
   Using Eq. (49), Eq. (56) can also be written as
      \beq
   v_\theta=-\frac{e\lambda_L}{m^*c}H_ce^{(r_0-r)/\lambda_L} .
   \eeq
   Note that London's equation for superconductors is
   \beq
   \vec{\nabla}\times\vec{v}=-\frac{e}{m*c}\vec{H}
   \eeq
   so in cylindrical coordinates
   \beq
   \frac{1}{r}\frac{\partial }{\partial r} (r v_\theta)=-\frac{e}{m*c}H=\frac{e}{m*c}H_c(e^{(r_0-r)/\lambda_L}-1)
   \eeq
   while Eq. (57) is, to lowest order in $\lambda_L/r$
       \beq
   \frac{1}{r}\frac{\partial }{\partial r} (r v_\theta)=\frac{e}{m*c}H_ce^{(r_0-r)/\lambda_L}
   \eeq
   so the velocity field of our perfect conductor definitely does $not$ satisfy London's equation.

\section{Effective mass reduction}
As discussed in section IV, the outward motion of electrons corresponds to both effective mass and bare mass flowing out,
while the outward motion of holes corresponds to effective mass flowing out and bare mass flowing in. So the process shown 
in Fig. 2 results in no bare mass flowing out,   but there is a net outflow of effective mass. 

For an electron in Bloch state $\vec{k}$ with band energy $\epsilon_k$, we define the effective mass $m^*_k$ by
\beq
\frac{1}{m^*_k}=\frac{1}{\hbar^2}\frac{\partial ^2\epsilon_k}{\partial k^2}
\eeq
assuming  there is no angular dependence for simplicity. For a given band we can define an effective mass density by
\beq
\rho_{m^*}=\int_{occ} \frac{d^3k}{4\pi^3} m^*_k
\eeq
where the integral is over the occupied (by electrons) states in the band. We can also of course define a bare mass density
\beq
\rho_{m}=\int_{occ} \frac{d^3k}{4\pi^3} m_e .
\eeq
Both $\rho_m$ and $\rho_{m^*}$ are zero for an empty band, for a full band $\rho_{m^*}=0$ and $\rho_m \neq 0$.
We can also define the associated mass and effective mass currents
\beq
\vec{j}_m=\int_{occ} \frac{d^3k}{4\pi^3} m_e \vec{v}_k
\eeq
\beq
\vec{j}_{m^*}=\int_{occ} \frac{d^3k}{4\pi^3} m^*_k \vec{v}_k
\eeq
with
\beq
\vec{v}_k=\frac{1}{\hbar} \frac{\partial \epsilon_k}{\partial \vec{k}} .
\eeq
Note  that the effective mass current density can also be written in the simple form
\beq
\vec{j}_m=\int_{occ} \frac{d^3k}{4\pi^3}( \frac{\partial \epsilon_k}{\partial \vec{v}_k}) .
\eeq
Both  real mass and effective mass currents satisfy continuity equations:
\bmath
\beq
\vec{\nabla}\cdot \vec{\j}_{m} +\frac{\partial \rho_{m}}{\partial t}=0 .
\eeq
\beq
\vec{\nabla}\cdot \vec{\j}_{m^*} +\frac{\partial \rho_{m^*}}{\partial t}=0
\eeq
\emath

When there is conduction in more than one band, the contributions from each band to the densities and currents simply add.
For the case under consideration here we have
\bmath
\beq
\vec{\nabla}\cdot \vec{\j}_{m,t}=0 
\eeq
\beq
\vec{\nabla}\cdot \vec{\j}_{m^*,t} = - \frac{\partial \rho_{m^*,t}}{\partial t}\neq 0
\eeq
\emath
where by the subindex $t$ (total) we mean the sum over both bands shown in Fig. 4.

We assume that the bands in Fig. 4 are respectively close to empty and close to full, so that the effective mass can be taken to be independent of
$k$ for the occupied states for the almost empty band and for the unoccupied states for the almost full band. 
Near the top of the band $m^*_k$ (Eq. (61)) is negative and we define the effective mass of holes near the top of the band as 
\beq
m_h^*=-m^*_k
\eeq
and for electrons near the bottom of the band 
\beq
m_e^*=+m^*_k .
\eeq

We have then for both bands,
denoted by $e$ and $h$
\bmath
\beq
\rho_{m^*}^e=\int_{occ} \frac{d^3k}{4\pi^3} m^*\equiv n_e m^*_e
\eeq
\beq
\rho_{m^*}^h=\int_{unocc} \frac{d^3k}{4\pi^3} m^*\equiv n_h m^*_h
\eeq
\emath
and furthermore assume $n_e=n_h=n_s$, so that no net outflow of mass occurs, with $n_s$ the superfluid density.

In the process shown in Fig. 2 there is a net outflow of $n_e$ electrons and $n_h$ holes, 
carrying out effective mass $m^*_e$ and $m^*_h$ respectively  per carrier, in the process where the magnetic field is expelled, i.e. in the process where the system goes from the normal to the superconducting state. This implies
\beq
\Delta \rho_{m^*}\equiv \rho_{m^*}^n-\rho_{m^*}^s=n_s(m^*_e+m^*_h)
\eeq
where the superscripts $n$, $s$ refer to normal and superconducting states. Therefore,
\beq
\rho_{m^*}^s= \rho_{m^*}^n - n_s(m^*_e+m^*_h) .
\eeq
which says that  the effective mass per carrier is lowered by ($m^*_e+m^*_h$)  when the system goes from the normal to the superconducting state
 and expels the magnetic field by expelling  electrons and   holes.
 
           \begin{figure} []
 \resizebox{8.5cm}{!}{\includegraphics[width=6cm]{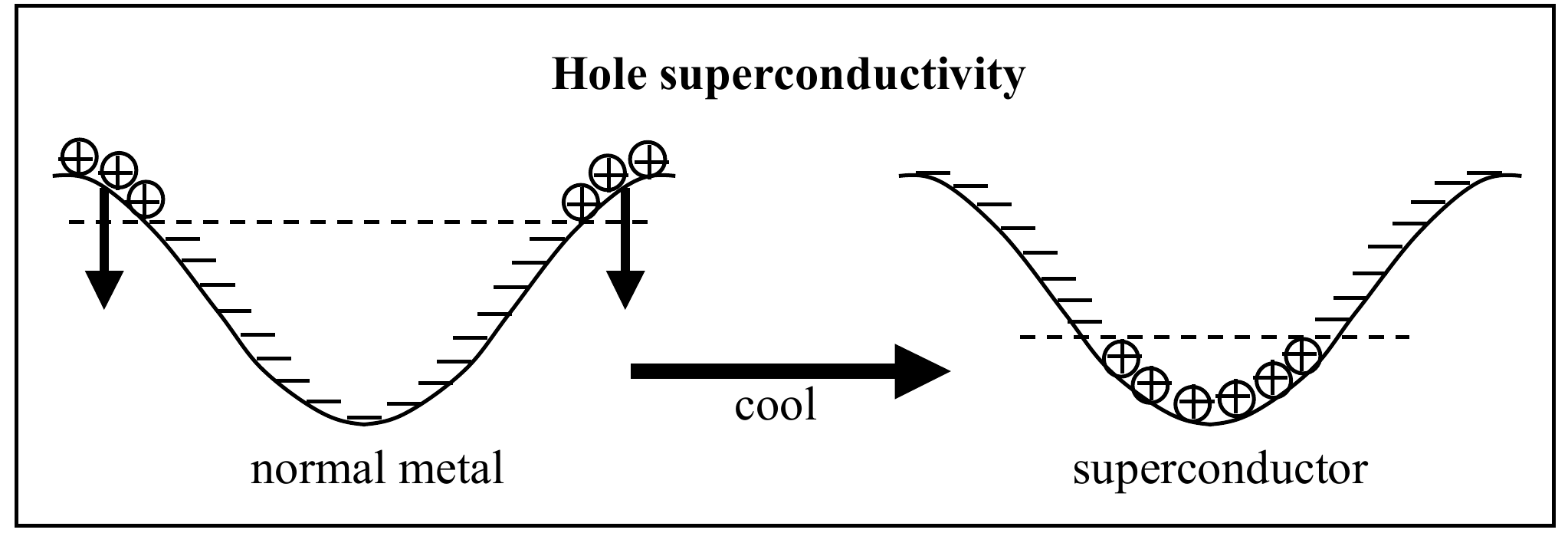}} 
 \caption {In the normal state of the metal, the band is almost full, with $n_h$ holes per unit volume that have 
 effective mass $m^*_h$.  As the metal becomes superconductor,
 the holes   move from the top  to the bottom of the band. This gives a reduction in the effective mass
 density of  $n_s(m^*e+m^*_h)$, with $m^*_e$ the effective mass of electron carriers at the bottom of the band.   }
 \label{figure1}
 \end{figure}

 Now recall that $n_s$, the superfluid density in the superconducting state, equals the density of charge carriers
 in the normal state, which is $n_e$ for a band close to empty and is   $n_h$ for a band close to   full.
 Therefore, our result Eq. (74)  is represented with what is shown in Fig. 6.   If the normal metal has a band that is almost full, with $n_h$ hole carriers with
effective mass $m^*_h$ (left panel), its effective mass density is
\beq
\rho_{m^*}^n=\int_{occ} \frac{d^3k}{4\pi^3} m^*_k=-\int_{unocc} \frac{d^3k}{4\pi^3} m^*_k=n_h m^*_h.
\eeq
The right panel of Fig. 6 depicts  $n_h$ empty states at the bottom of the band. The effective mass density 
for that situation is
\beq
\rho_{m^*}^s=\int_{occ} \frac{d^3k}{4\pi^3} m^*_k=-\int_{unocc} \frac{d^3k}{4\pi^3} m^*_k=-n_h m^*_e  .
\eeq
Therefore,  Eq. (74) is satisfied.

As seen in Fig. (6), the physics we are finding $requires$ that in the normal state the charge
carriers are holes, as proposed in the theory of hole superconductivity \cite{holesc}. 
Furthermore, Fig. (6) indicates  that  when the system goes from normal metal to superconductor the holes near the top
of the band `Bose condense' into states at the bottom of the band.  We discuss this further in Sect. X.

\section{angular momentum conservation}
The important issue of angular momentum conservation needs to be addressed. In the process shown in Fig. 2, 
the final state has angular momentum given by \cite{momentum}
\beq
L=(2\pi R\lambda_Lhn_s)m_e v_s R=\frac{m_ec}{2e}R^2 h H_c .
\eeq
How did electrons acquire this angular momentum, and how is
angular momentum conserved?

For the discussion in Sect. V we assumed initial conditions so that only electrons have azimuthal velocity.
However, let us consider first the simpler situation where the initial velocity is zero for both negative and positive charges.

As the perfectly conducting fluid starts moving outward, after a time $t_0\sim \lambda_L/\dot{r}_0$ negative and positive charges near the inner boundary have acquired equal and opposite azimuthal velocities due to the
action of the magnetic Lorentz force, giving rise to the azimuthal current density Eq. (18) as the sum of both contributions.
The total angular momentum is thus zero. As the fluid moves out, both negative and positive charges increase their angular momentum,
and at the end they both attain half the value Eq. (41), 
but their sum remains zero at all times. Thus conservation of angular momentum follows naturally in this scenario. 
However, it still needs to be explained how the charges increase their angular momentum as the fluid moves out, given that 
we said in Sect. V that the electric and magnetic forces in the azimuthal direction are balanced for both negative and positive charges (Fig. 5).

The reason is, the treatment given in Sect. V was approximate, valid to lowest order $\lambda_L/r$. Recall also that we found for example that Eq. (6) was satisfied only to
lowest order in $\lambda_L/r$. An exact treatment is more complicated
and requires the use Bessel functions. One finds that in fact the electric and magnetic forces are not exactly balanced, the electric force is
slightly larger, providing the necessary torque so that the azimuthal velocity does not slow down but rather stays constant as the fluid
moves out, thus imparting the increasing angular momentum to the currents.

Going back to the scenario where only electrons have azimuthal velocity shown in figure 5, it would require a very artificial initial condition:
that  both electrons and holes  initially have azimuthal velocity in counterclockwise direction given by half the 
value Eq. (17) , so that   in the  outward motion   the Lorentz force causes the holes to stop and the electrons to
double their initial velocity.  This is not what we say happens in the Meissner effect.

In the next section we will discuss what really happens in the Meissner effect
according to the theory of hole superconductivity \cite{holesc}. But it should be clear from the discussion here and in Sect. V that
the essential physics of magnetic field expulsion follows simply from these magnetohydrodynamic considerations.

\section{what really happens}

The process depicted in Fig. 2 shows the essential physics of   what we argue  is required to expel the 
magnetic field in the normal-superconductor transition. But it is only a caricature of what really happens, it cannot be the
reality. In particular, holes flowing {\it out of} the outer boundary of the annulus in Fig. 2  implies that electrons
are flowing $into$ the outer boundary of the annulus. But where did those electrons come from?

The theory of hole superconductivity \cite{holesc} provides the answer. 
We review the physics here, discussed in earlier references \cite{sm,bohr,ondyn,disapp,revers,momentum,whyholes,entropy}.
It requires that the normal state charge carriers are in a band that is almost full, with hole concentration 
$n_s$ that will become the superfluid density.

First, the theory predicts that when electrons condense into the superconducting state their orbits expand from a microscopic radius to
radius $2\lambda_L$ \cite{sm}. The radius is determined by quantization of angular momentum \cite{bohr}.

This orbit expansion is equivalent to an $outflow$ of the electron negative charge a distance $\lambda_L$.
To preserve charge neutrality, an $inflow$ of normal electrons has to occur over that distance. These normal
electrons are in a band that is almost full, so they represent  an inflow of negative effective mass carriers,
or equivalently an outflow of holes over the same distance. The process
is depicted in Fig. 7.

          \begin{figure} []
 \resizebox{6.5cm}{!}{\includegraphics[width=6cm]{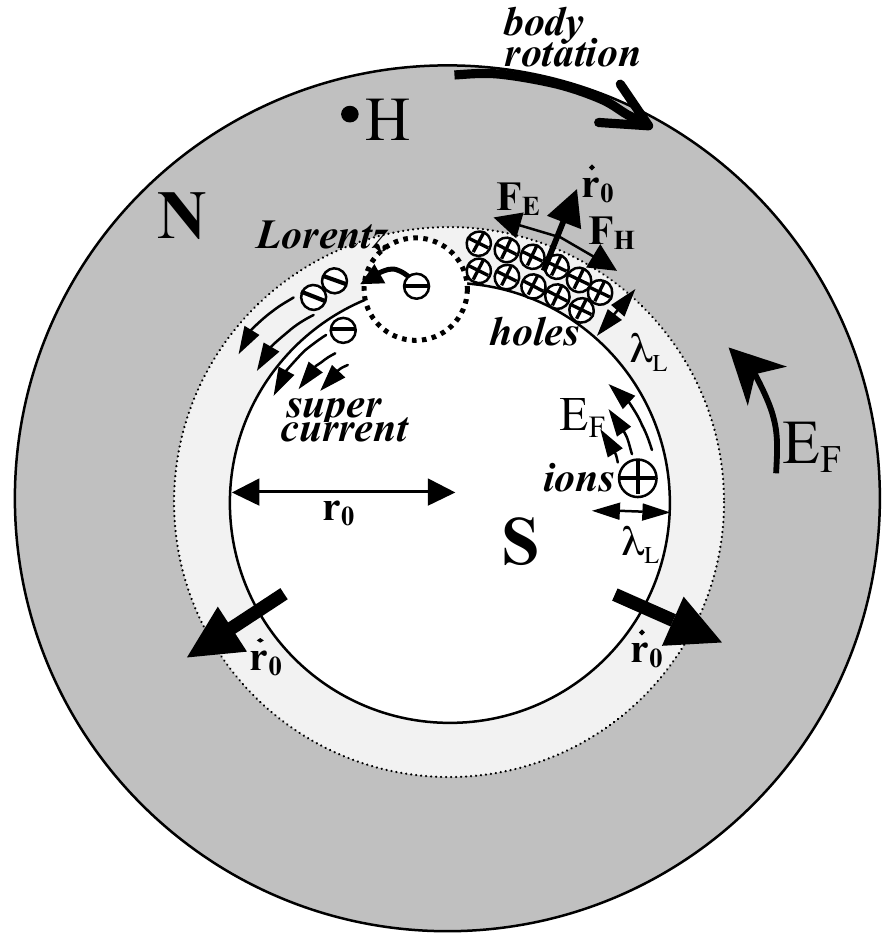}} 
 \caption { Meissner effect according to the theory of hole superconductivity.  
 As normal electrons become superconducting their orbits (dotted circle) expand, and the resulting Lorentz force propels the supercurrent. An outflow of hole carriers moving in the same direction as the phase boundary restores charge neutrality and transfers momentum to the body as a whole to make it rotate clockwise, without any scattering processes.}
 \label{figure1}
 \end{figure} 

The electric and magnetic Lorentz forces acting on the holes are balanced as shown in Fig. 7, just as we showed
in Fig. 5 in our `caricature' process. The holes move out radially at speed $\dot{r}_0$, the speed of motion of the phase
boundary, with no azimuthal velocity.

On the electrons, electric and magnetic forces are not balanced. We assume the orbit expansion occurs at great speed
(much larger than $\dot{r}_0$). In expanding the orbit to radius $2\lambda_L$   the electrons acquire azimuthal (counterclockwise) velocity 
\beq
v_s=-\frac{e\lambda_L}{m^*c}H_c
\eeq
driven by the
magnetic Lorentz force, with the electric Faraday force in the opposite direction having negligible effect \cite{momentum}.

The Faraday electric field is slightly different than in our simple model of Sect. V, it is given by
\bmath
\beq
\vec{E}=\frac{\dot{r}_0}{c}H_c e^{(r-r_0)/\lambda_L} \hat{\theta}
\eeq
for $r\leq r_0$, and
\beq
\vec{E}=\frac{\dot{r}_0}{c}\frac{r_0}{r}H_c  \hat{\theta}
\eeq
\emath
for $r\geq r_0$.  The azimuthal speed of electrons is
\beq
\vec{v}_s(r)=-\frac{e\lambda_L}{m^*c}H_c e^{(r-r_0)/\lambda_L} \hat{\theta}
\eeq
for $r\leq r_0$ and zero for $r>r_0$ (except for a small normal current induced by $E$ \cite{momentum}). 
Note that the speed increases  with $r$, in contrast to the situation in
Sect. V where it decreases (see Fig. 5). As the phase boundary moves further out, the Faraday electric field slows down the
azimuthal velocity Eq. (80) as the given point $r$ gets further away from the phase boundary, and both 
$\vec{E}$ and $\vec{v}_s$ go to zero in the deep interior \cite{momentum}.

           \begin{figure} []
 \resizebox{6.5cm}{!}{\includegraphics[width=6cm]{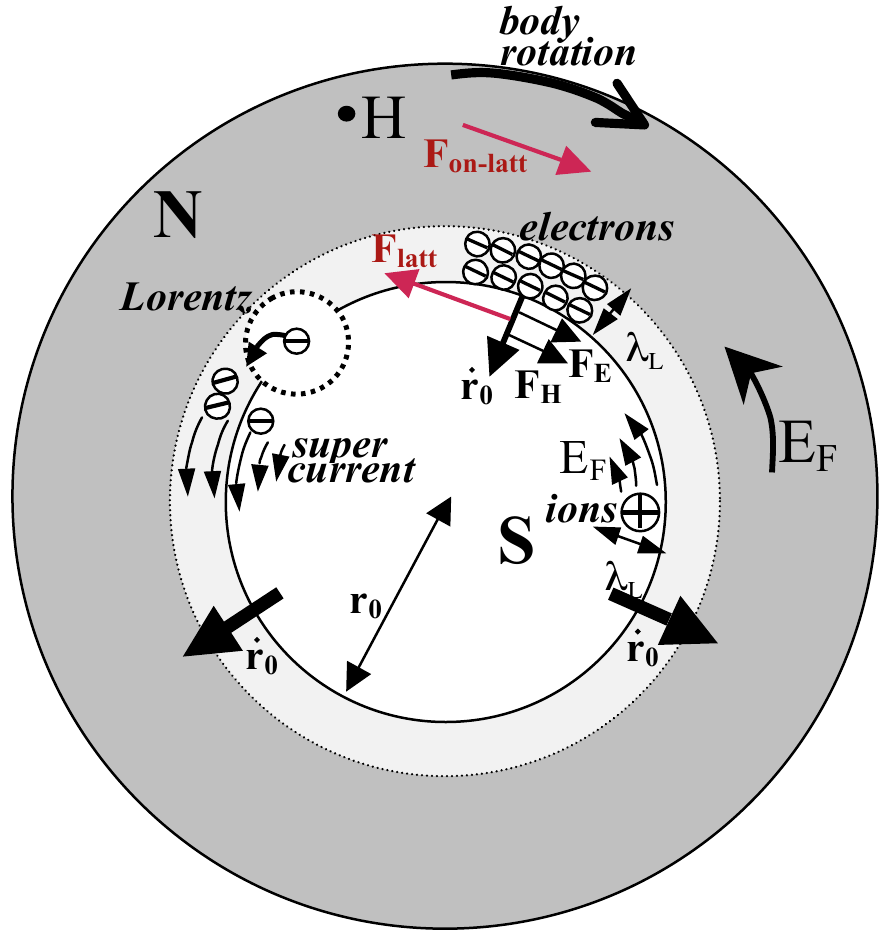}} 
 \caption {Figure 7 redrawn replacing the outflowing holes by inflowing electrons.   The electric and magnetic forces on inflowing electrons $F_E$ and $F_H$ point in the same direction. Since the motion is radial, this implies that another force must exist, $F_{latt}$, exerted by the periodic potential of the ions on the charge carriers. By NewtonÕs third law, an equal and opposite force is exerted by the charge carriers on the ions, $F_{onlatt}$, that makes the body rotate.}
 \label{figure1}
 \end{figure}

Fig. 8 shows the same process as fig. 7 with the outflowing holes replaced by inflowing electrons. 
It clarifies the important issue of angular momentum balance \cite{momentum,revers}. As the electrons in the expanding orbits acquire their azimuthal speed
their increasing angular momentum has to be compensated by the body as a whole acquiring angular momentum in opposite
direction. This happens through the backflow of electrons with negative effective mass, i.e. outflow of holes. 
The lattice exerts an azimuthal force $F_{latt}$ on these electrons,
and in turn these electrons exert a force on the lattice $F_{on-latt}$ that transfers angular momentum to the body
without any scattering processes that would lead to irreversibility. 
It is essential that the normal state charge carriers are holes. This is 
a key issue explained in detail in the references \cite{momentum,disapp,whyholes}.

 In summary, `what really happens' is not exactly the same but very similar in spirit to the `caricature' process
 shown in Fig. 2 and discussed in Sect. V, that could be understood simply using (almost) purely classical
 concepts. The difference here is that it is not the same electrons and holes that move continuously
 out, as in Fig. 2. Rather, electrons right outside the phase boundary move  out  a distance
 $\lambda_L$ when they enter the superconducting state, and normal electrons from a distance up to $\lambda_L$
 outside the phase boundary move  in. The region inside the phase boundary ends up in the superconducting
 state, having expelled $n_s m^*$ and absorbed $-n_s m^*$ effective mass density in the process, or equivalently
 having lowered its effective mass density by $2 m^*$ per normal state carrier, as we discussed in Sect. VII.

\section{the physics of hole superconductivity}
We have described  the motion of magnetic field lines when a normal metal goes superconducting
using concepts used in describing the magnetohydrodynamics of conducting fluids, and in particular Alfven's theorem.
 Let us recapitulate our reasoning.

Starting from the observation that perfectly conducting fluids drag magnetic field lines with them when they move,
we suggested that the moving field lines in the Meissner effect are dragged by a perfectly conducting fluid. 
We argued that this fluid has to be both charge-neutral and mass-neutral in order to not generate charge nor mass imbalance.
We concluded that in order for this to happen it is necessary that the system expels the same concentration of electrons and holes.

We found that this implies that when the system goes from normal to superconducting and expels a magnetic field it also expels effective mass,
 so the  effective mass in the system is reduced in going from the normal to the superconducting state. The amount of effective mass reduction per superfluid carrier was found
to be independent of the magnitude of the magnetic field expelled. This then leads us to the general conclusion that when a system goes
superconducting the carriers lower their effective mass, whether or not a magnetic field is present.

It is interesting to note that back in 1950 John
Bardeen proposed a model of superconductivity
which had as an essential ingredient a reduction of
the carriers' effective mass upon entering the superconducting
state \cite{bardeenmstar}. However the model did
not include the pairing concept, and in the
subsequent BCS theory   the effective mass
reduction concept was not incorporated.

Within the theory of hole superconductivity   \cite{holesc} the interaction that gives rise  to pairing is a  correlated hopping term $\Delta t$  in the effective Hamiltonian that increases
the   mobility of  carriers when they pair \cite{hole1,bondch}, or in other words decreases their effective mass.
Superconductivity is driven by lowering of
kinetic energy or equivalently by effective mass reduction. There is a  lowering of the effective mass of Cooper pairs
relative to the effective mass of the normal carriers   \cite{stanford,bose,mass}, and this  gives rise to a London penetration depth that is smaller than expected from the normal state effective mass \cite{londonours}.
This in turn leads to an apparent violation \cite{apparent,color} of the low frequency optical conductivity sum rule (Ferrell-Glover-Tinkham sum rule)
\cite{99percent,optical}
that was detected experimentally in several high $T_c$ superconductors  years after first predicted
\cite{marel1,marel2,basov,keimer}.

More fundamentally the theory predicts that carriers `undress' in the transition from the normal to the superconducting state \cite{color,undressing,undressing3}, both lowering their effective mass and increasing their
quasiparticle weight \cite{undressing4}. In a many-body system, the quasiparticle weight is inversely proportional to the effective mass,
a highly dressed particle has both large effective mass and small quasiparticle weight and vice versa \cite{mattuck}. 
Clear experimental evidence for increase in the quasiparticle weight upon onset of superconductivity 
has been found in the cuprate superconductors \cite{ding}.

Even more fundamentally, the theory predicts that carriers undress from $both$ the electron-electron
interaction and  the electron-ion interaction \cite{ehasym0,atom,undrelion,holeelec2}. 
In the normal state of the system when the band is almost full, i.e. when the normal state carriers are holes, carriers are `dressed' by the electron-ion interaction causing the electrons at the Fermi energy to have negative rather than positive effective mass. When the system goes superconducting,
experiments and theory clearly show that the $n_s$ superfluid carriers are `undressed' from the electron-ion interaction
because they behave as electrons with negative charge \cite{londonmom,gyro,bernoulli}. For example, a rotating superconductor shows always a
magnetic field in direction parallel, never antiparallel, to its angular velocity \cite{londonmom}.

The latter was understood to reflect the fact that the wavelength of carriers expands when they go from normal
to superconducting \cite{holeelec2}. Normal state carriers at the top of the  band interact strongly with the discrete ionic potential and when they go superconducting
and their wavelength expands    they no longer `see' the discrete ionic potential, hence have `undressed' from it and behave
as electrons rather than holes. More specifically, the wavelength expansion was found to result from   electronic   orbits expanding 
from a microscopic radius to radius $2\lambda_L$ \cite{sm,missing}  in the   transition.
All of this led us to conclude that `holes turn into electrons' in the  normal to superconducting transition \cite{reduc,ehasym}.
Based on this physics, supported by  quantitative calculations, the picture shown in Fig. 6 was proposed in 2010,
Fig. 6 of Ref. 
  \cite{holeelec3}. What this means for the wavefunction of the carriers is what is shown in Fig. 9 \cite{holeelec3}.
  In the normal state, carriers at the Fermi energy are in `antibonding states', with highly oscillating wavefunction
  and  high kinetic energy, while in the superconducting state they adopt the same wavefunction that
  electrons have near the bottom of the band in the normal state, i.e. bonding states, with smooth 
  wavefunction and low kinetic energy. Figure 6 expresses this fact.

                     \begin{figure} []
 \resizebox{8.5cm}{!}{\includegraphics[width=6cm]{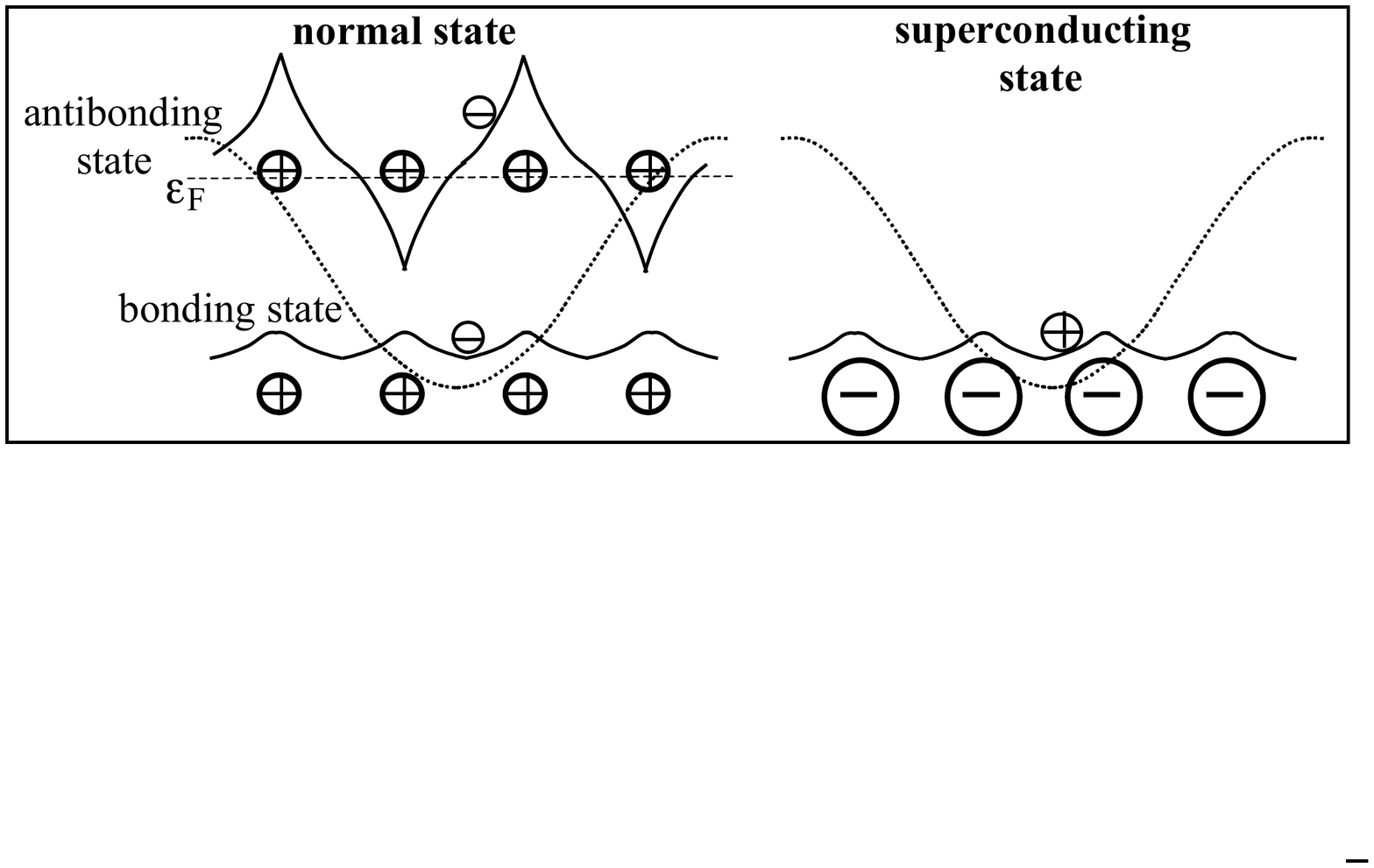}} 
 \caption {When a band is nearly full, carriers at the Fermi energy (indicated by $\epsilon_F$) are in antibonding states
  (left panel), with
 highly oscillating wavefunction and high kinetic energy. Carriers near the bottom of the band are in bonding states, with smooth
 wavefunction and low kinetic energy. According to the finding in Fig. 6, when a system becomes superconducting and
 expels electrons and holes, the wavefunction for the superconducting carriers becomes as shown in the right
 panel of the figure, a bonding state. }
 \label{figure1}
 \end{figure} 

In this paper we have independently `rediscovered' Figs. 6 and  9 by finding  that the Meissner effect requires normal state carriers of density $n_s$
to lower their effective mass by $(m^*_e+m^*_h)$ as they go superconducting, or equivalently that they change their
effective mass from $-m^*_h$ to $m^*_e$. This requires that the initial state has a band that is almost full,
with hole carriers of mass $m^*_h$, 
and that in going superconducting the holes move from the top to the bottom of the band as shown in
Fig. 6. 

The $requirement$ that normal state carriers in a metal that can become a superconductor are
holes rather than electrons \cite{whyholes} follows directly from this physics. There would be  no way for carriers 
to  lower their effective mass by $(m^*_e+m^*_h)$   starting with a normal state with electron
carriers of effective mass $m_e$.

\section{dissipation}
In the process shown in Fig. 2, magnetic field lines are carried out by a perfectly conducting fluid, 
and no dissipation is associated with that motion \cite{davidson}. However, Joule heat is still generated due to the motion of magnetic field lines
outside the region occupied by the perfectly conducting fluid. Let us calculate that.
At a given time, the perfectly conducting fluid in Fig. 2 occupies an  annulus $r_0<r<r_1$, with
\beq
r_1^2=r_0^2+2R\lambda_L .
\eeq
The electric field for $r>r_1$ is
\beq
E(r)=\frac{r_0}{r}\frac{\dot{r}_0}{c}H_c
\eeq
and the Joule  heat dissipated per unit volume per unit time in the region $r_1<r<R$ is
\beq
\frac{\partial w}{\partial t}=\sigma \frac{\dot{r}_0^2}{c^2}\frac{r_0^2}{r^2}H_c^2 .
\eeq
The total heat per unit time dissipated in the region $r_1<r<R$ is
\beq
\frac{\partial W}{\partial t  }=\int_{r_1}^R  d^3r \frac{\partial w}{\partial t}= 2\pi h H_c^2\frac{\sigma}{c^2}\dot{r}_0^2 r_0^2\int_{r_1}^R dr \frac{1}{r}
\eeq
and integrated over time
\beq
W=2\pi h H_c^2 \frac{\sigma}{c^2}\dot{r}_0 \int _0^{R-\lambda_L} dr_0 r_0^2 ln(R/r_1)
\eeq
assuming for simplicity that $\dot{r}_0$ is time-independent. 

Instead, let us assume that the magnetic field gets expelled through some unknown quantum mechanism that does not  involve
the motion of a perfectly conducting fluid, as in BCS. The same equation (84) applies with $r_0$ replacing $r_1$, hence
\beq
\frac{\partial W_0}{\partial t  }=2\pi h H_c^2\frac{\sigma}{c^2}\dot{r}_0^2 r_0^2\int_{r_0}^R dr \frac{1}{r}
\eeq
and the total Joule heat dissipated is
\beq
W_0=2\pi h H_c^2 \frac{\sigma}{c^2}\dot{r}_0 \int _0^{R-\lambda_L} dr_0 r_0^2 ln(R/r_0) .
\eeq
Carrying out these integrals we find, to lowest order in $\lambda_L/R$
\beq
W_0=\frac{2\pi}{9}hR^3H_c^2 \frac{\sigma}{c^2}\dot{r}_0
\eeq
and 
\beq
W=W_0-\Delta W
\eeq
with
\beq
\Delta W=12\frac{\lambda_L}{R}W
\eeq
so the Joule heat dissipated is less when the process occurs through motion of a perfectly conducting fluid as
in Fig. 2.

We argue that something similar occurs for the Meissner effect. If we assume that the expulsion of magnetic field
occurs without any radial motion of charge, as in BCS, the Joule heat dissipated will be the same as Eq. (88), i.e.
\beq
Q_J^0=\frac{2\pi}{9}hR^3H_c^2 \frac{\sigma}{c^2}\dot{r}_0 .
\eeq
Instead, if the process occurs through the flow and backflow of electrons and holes discussed in Sect. IX, 
the Joule heat will be smaller because in the region $r_0<r<r_0+\lambda_L$ no dissipation occurs \cite{revers},
so instead of Eq. (86) we have 
\beq
\frac{\partial Q_J}{\partial t  }=2\pi h H_c^2\frac{\sigma}{c^2}\dot{r}_0^2 r_0^2\int_{r_0+\lambda_L}^R dr \frac{1}{r}
\eeq
and we obtain
\beq
Q_J=Q_J^0-\Delta Q_J
\eeq
\beq
\Delta Q_J=\frac{3}{2}(\frac{\lambda_L}{R})  Q_J^0  .
\eeq

              \begin{figure} []
 \resizebox{7.5cm}{!}{\includegraphics[width=6cm]{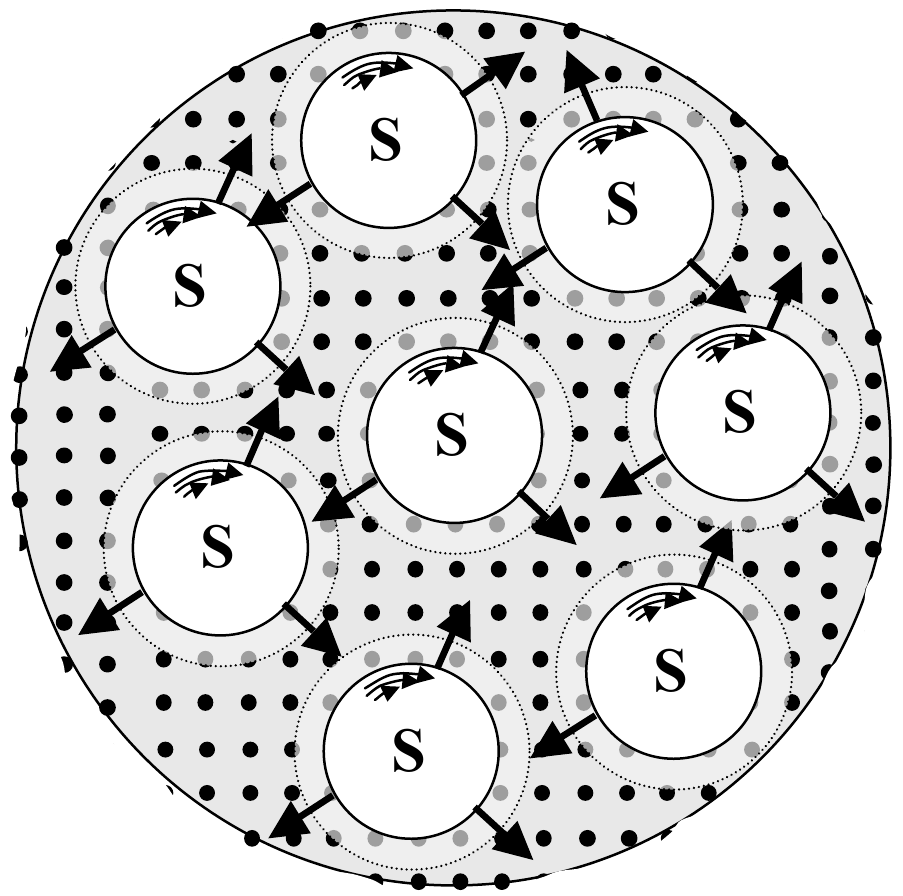}} 
 \caption {Expulsion of magnetic field through nucleation of several superconducting domains that expand
 and merge.  In the annuli  of thickness $\lambda_L$ beyond the surface of each domain, indicated by the
 dotted circles, no Joule dissipation occurs in the process discussed in Sect. IX. }
 \label{figure1}
 \end{figure} 
 
The correction Eq. (94) is small if $\lambda_L<<R$. However note that we are assuming that
the flux expulsion occurs through the expansion of a single domain as shown in Fig. 7.
Consider instead the more realistic scenario where the superconducting phase nucleates
at several different point simultaneously, creating several domains that expand expelling magnetic
field, as shown in Fig. 10. Now we need to exclude a region of thickness $\lambda_L$ around
each domain where dissipation will not take place for the calculation of $Q_J$. 
This corresponds roughly to replacing $R$ by $R/N$, with $N$ the number of domains, in Eq. (94).
So we conclude that if the transition occurs through nucleation of many domains the Joule heat dissipated as
the magnetic flux is expelled will be drastically reduced. 
Instead, if dissipation occurs without accompanying fluid motion, it will be the same whether a single   domain
or many domains are involved. 

It should be possible to observe this experimentally. In the absence of the radial charge flow  predicted in our
theory, it would be difficult to explain why the Joule heat generated would be any different if the transition 
and flux expulsion occurs
via a single domain that expands, or via many different domains, for a given speed of motion of field lines
through the dissipating regions. Experimentally it should be possible to realize
the different domain scenarios by setting up appropriate  temperature or magnetic field gradients.

\section{discussion}

None of the physics discussed in this paper is part of  the conventional theory of superconductivity \cite{tinkham}. 
About  the Meissner effect  the conventional theory  simply says that magnetic field
lines move out because the superconducting state with magnetic field excluded has lower energy than the
normal state with magnetic field inside.   The dynamics of the Meissner effect and  of  the related effect of magnetic field generation when a  rotating
normal metal becomes superconducting  (London field) \cite{londonmom,inertia2}  have not been explained within
the conventional theory. 
The BCS `proof' of the Meissner effect \cite{tinkham} starts with the system in the superconducting state and applies a magnetic
field as a small perturbation, which is $not$ the physics of the Meissner effect \cite{ondyn}.

It is important to remember  that the   laws of classical physics that we used in this paper always act, whether or not
`quantum mechanics' also plays a role. Specifically, in addition to explaining how `quantum mechanics' causes
magnetic field lines to be expelled, the conventional theory has to explain how angular momentum is conserved
and how  the process  overcomes the laws of classical physics that say
that magnetic field lines have great difficulty in moving through conducting fluids, the more so the more conducting
the fluid is, and that   energy is dissipated in the process, and entropy is generated. The normal-superconductor transition in a magnetic field is a reversible phase transformation that  occurs without
entropy generation in an ideal situation. Entropy is not generated when magnetic field lines move 
following the motion of a perfectly conducting fluid,   while entropy is generated when magnetic field lines
cut across a conducting fluid, whether or not quantum mechanics plays a role. Within our theory, entropy is not generated locally around the
phase boundary when the phase boundary is displaced, while it would be within the conventional theory.
Based on this we have proposed that Alfven waves \cite{alfventheorem} should propagate along normal-superconductor phase boundaries
if our theory is valid \cite{alfvenwaves} and not if the conventional theory is valid.
Elsewhere we argue that Alfven waves should also propagate near the surface of 
a superconductor in the presence of a magnetic field   \cite{alfvensound}.

 Within the conventional theory the only thing that flows out when a system goes
from normal to superconducting and expels a magnetic field is `phase coherence'. Nobody has explained even qualitatively  how this abstract concept explains the
physical processes that take place, that at face value appear to violate fundamental laws of physics, namely  the law of inertia, Faraday's law, conservation of angular momentum 
and conservation of entropy in reversible processes \cite{ondyn,momentum,revers,entropy}.

Instead, in this  paper we have argued that magnetohydrodynamics strongly suggests that the Meissner effect in superconductors is associated
with outflow of a perfectly conducting fluid in the normal-superconductor transition. 
That  this perfectly conducting fluid needs to be composed
of electrons and holes, to preserve charge neutrality and mass homogeneity. That electrons becoming superconducting
flow out, and there is a backflow of normal antibonding electrons equivalent to an outflow of normal holes, and that momentum is conserved by holes
transferring it to the body as a whole. The  process as we describe it  is reversible, as required by thermodynamics, 
and satisfies the fundamental laws of physics. In this paper we described the process
 in more detail than in earlier work \cite{ondyn,momentum}
and unexpectedly found that it    leads to a lowering of effective mass in going from normal to superconducting,
in unexpected agreement with what the theory of hole superconductivity has predicted for the last 30 years and
was found experimentally in some  high temperature superconductors \cite{marel1,marel2,basov,keimer,ding}.
We also found in this paper that the process requires the normal state carriers to be hole-like
for yet another reason that adds to  the many other reasons found in earlier work \cite{holesc}, 
  and in contrast to the conventional theory of superconductivity
that is electron-hole symmetric  \cite{tinkham}. Macroscopic phase coherence also follows naturally
from this physics \cite{holeelec2,reduc}.

The `population inversion' that we found in Fig. 6 is reminiscent of what occurs in laser physics. It is interesting
that in both realms it is associated with the establishment of macroscopic quantum phase coherence.
In ref. \cite{holeelec3} we presented many other reasons for why the scenario shown in Figs. 6 and 9 captures the
essence of superconductivity. We argue that the fact that in this paper we have `rediscovered' Figs. 6 and
9 from an entirely different argument strongly supports the validity of this theoretical framework to describe
the real world, as opposed to the world  of `model Hamiltonians' \cite{model1,model2,model3,model4}.
It represents a radical departure from the conventional theory of superconductivity, where it is 
assumed that carriers establish correlations between each other when they become superconducting but do not change
their intrinsic character, i.e. their wavefunction. Instead, in our scenario carriers change their
most essential characteristics, their quasiparticle weight and effective mass, because their wavefunction changes,
through the complete redistribution  of energy level occupation depicted in Fig. 6.

As shown in ref. \cite{holeelec3} and earlier   papers, this physics also leads to a slight charge inhomogeneity in the ground state of superconductors
\cite{chargeexp},
with more negative charge near the surface and more positive charge in the interior, and to macroscopic zero point motion in the form
of a spin current flowing near the surface of superconducting bodies in the absence of 
applied fields \cite{atom,spinc,electrospin}. 

                   \begin{figure} []
 \resizebox{6.5cm}{!}{\includegraphics[width=6cm]{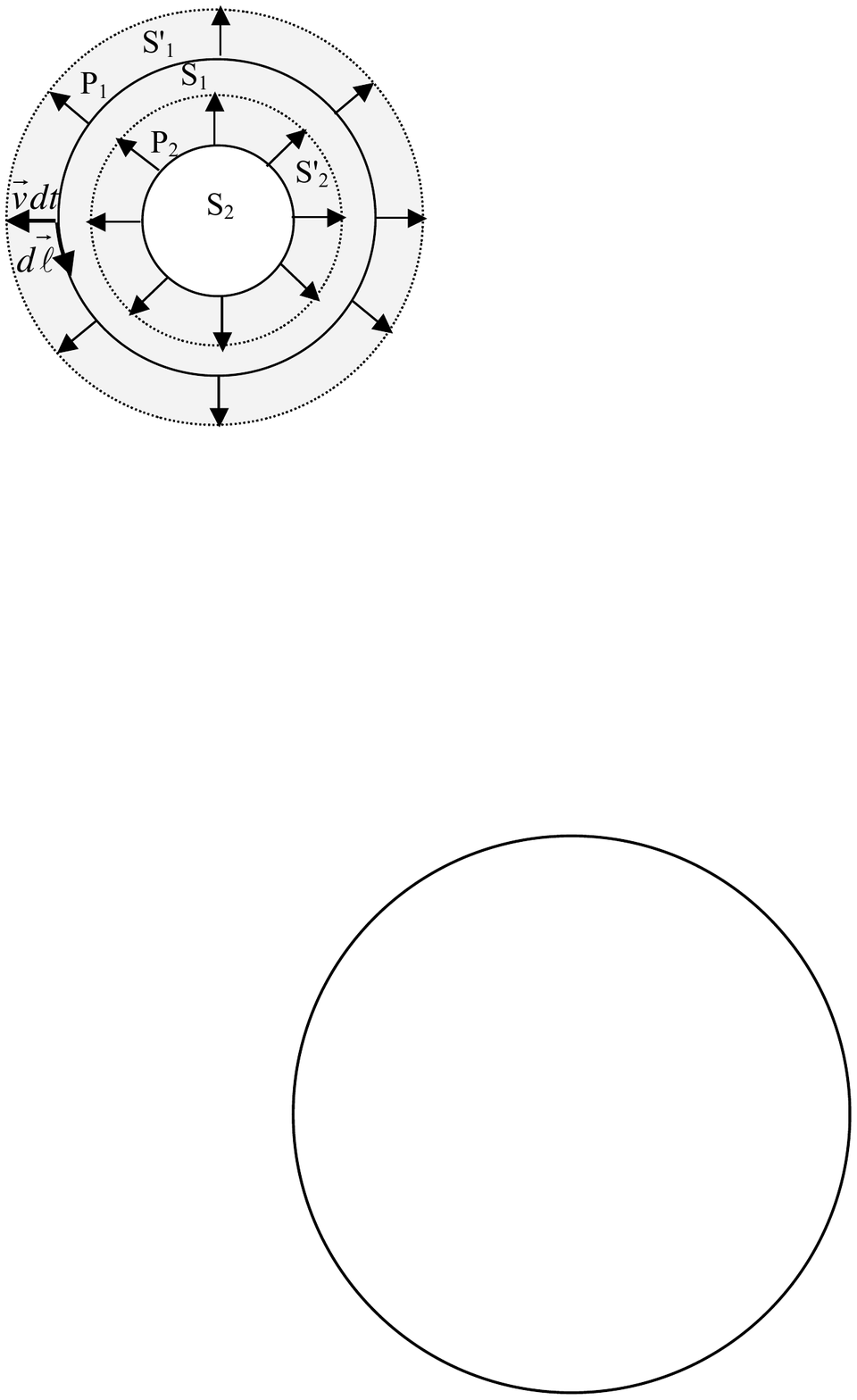}} 
 \caption { Outward motion of perfectly conducting region. $P_1$ and $P_2$ are the outer and inner boundaries of the perfectly conducting region, indicated by full circes. $dS_1$ ($dS_2$)  is the area bounded by the outer (inner) full circle and the outer (inner) dotted circle.}
 \label{figure1}
 \end{figure} 
                    \begin{figure} [t]
 \resizebox{8.5cm}{!}{\includegraphics[width=6cm]{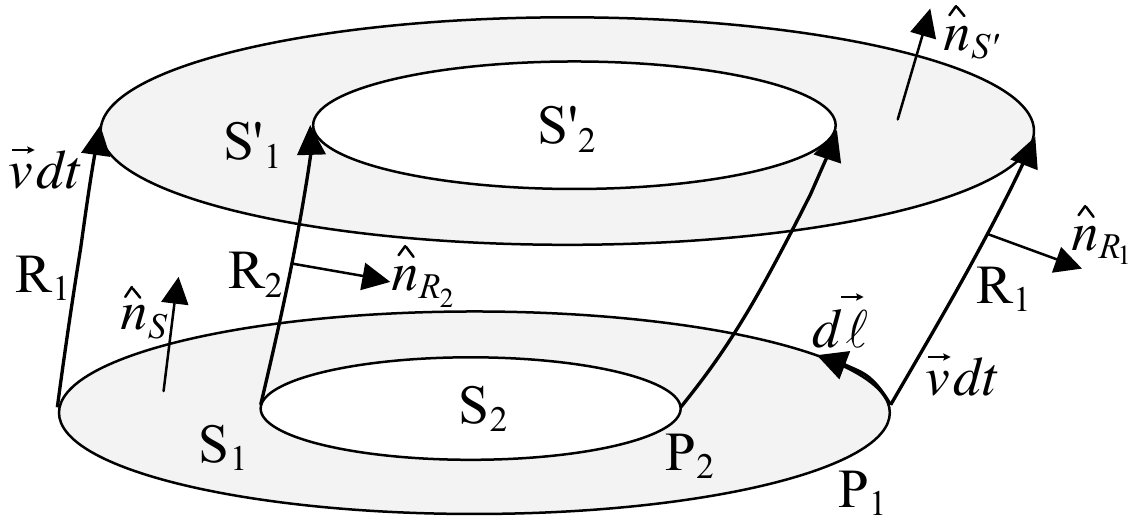}} 
 \caption { Alfven's theorem for a more general geometry than in Fig. 11. Surfaces $S_1$ and $S_2$ bounded by perimeters 
 $P_1$ and $P_2$ evolve
 to $S'_1$ and $S'_2$ in time $dt$ following the motion of the fluid. $R_1$ and $R_2$ denote the ribbons connecting the surfaces under time
 evolution. The magnetic flux through the grey region stays the same for all times.  }
 \label{figure1}
 \end{figure}

 We suggest  that a valid microscopic theory of superconductivity has to be consistent with these findings.
 The currently accepted conventional theory of superconductivity \cite{tinkham}  certainly  is not \cite{validity}.
\newline
\appendix 
  \section{Alfven's theorem}

We  prove Alfven's theorem for the case of interest  in this paper shown in Fig. 2. Figure 11 shows a slice of our cylinder 
  perpendicular to its axis evolving in time. The full lines show the  outer and inner perimeters of the perfectly conducting fluid at time $t$, denoted by $P_1$, $P_2$, which are boundaries to
  the surfaces $S_1$ and $S_2$ which become $S_1'$ and $S_2'$ at time $t+dt$, bounded by the dotted lines.
  We want to show that the magnetic flux through $S_1$ and $S_1'$ are the same, as well as that through 
  $S_2$ and $S_2'$. This then implies that the flux through the annulus doesn't change, and also  that the flux through
  the central region inside the annulus remains zero.

  The change in magnetic  flux through one of these surfaces, $d\phi_i$  ($i=1$ or $2$)  is
  \beq
  d\phi_i =\int_{S_i'}\vec{B}(\vec{r},t+dt)\cdot d\vec{a}-\int_{S_i}\vec{B}(\vec{r},t)\cdot d\vec{a}
  \eeq
  Using 
  \beq
  \vec{B}(\vec{r},t+dt)=\vec{B}(\vec{r},t)+\frac{\partial B(\vec{r},t)}{\partial t} dt
  \eeq
  we have
    \beq
  d\phi_i =\int_{dS_i}\vec{B}(\vec{r},t)\cdot d\vec{a} + dt\int_{S_i}\frac{\partial B(\vec{r},t)}{\partial t}\cdot d\vec{a}
  \eeq
  with $dS_i=S_i'-S_i$. The differential of area is
  \beq
  d\vec{a}=\vec{v}\times d\vec{\ell}
  \eeq
  where $\vec{v}$ is the velocity of the fluid. Using Eq. (A4) for the first integral and Eq. (6) for the second integral in Eq. (A3) we obtain
  \beq
   d\phi_i =dt \int_{P_i}\vec{B}(\vec{r},t)\cdot (\vec{v}\times d\vec{\ell})+dt\int_{S_i} \vec{\nabla}\times(\vec{v}\times\vec{B}) \cdot d\vec{a}
   \eeq
   and permuting factors in the first term and using Stokes' theorem in the second term
    \beq
   d\phi_i =dt \int_{P_i}(\vec{B} \times \vec{v}) \cdot d\vec{\ell}+dt\int_{P_i}(\vec{v}\times\vec{B}) \cdot d\vec{\ell}=0 .
   \eeq
  This proves that magnetic field lines don't cross neither the outer nor the inner boundary as the annulus of perfectly
  conducting fluid moves outward. Magnetic field lines are frozen into the annulus  and move out with it, pushing out magnetic field lines outside and 
  leaving the interior
  field free.

  For a more general geometry where the fluid velocity is not parallel to the area being considered, as shown in Fig. 12  the proof is only slightly more complicated. 
  In addition to the flux through the surfaces $S_1$ and $S_2$ we need to consider also the flux through the ribbons $R_1$ and $R_2$ shown
  in Fig. 11. It can be shown that here also the flux through the multiply connected surface bounded by $P_1$ and $P_2$ is invariant under
  time evolution, and the flux in the interior of $P_2$ remains zero at all times.

\acknowledgements
The author is grateful to F. Marsiglio for helpful comments.

\end{document}